\documentstyle[amsfonts,prl,aps,multicol]{revtex}
\input psfig
\pssilent
\begin{document}

\title{Absence of Singularity in Loop Quantum Cosmology}
\author{Martin Bojowald\cite{Email}}
\address{Center for Gravitational Physics and Geometry, The
  Pennsylvania State University,\\ 104 Davey Lab, University Park, PA
  16802, USA}

\maketitle

\begin{abstract}
  It is shown that the cosmological singularity in isotropic
  minisuperspaces is naturally removed by quantum geometry. Already at
  the kinematical level, this is indicated by the fact that the
  inverse scale factor is represented by a bounded operator even
  though the classical quantity diverges at the initial 
  singularity. The full demonstation comes from an analysis of 
  quantum dynamics. Because of quantum geometry, the quantum evolution 
  occurs in {\em discrete\/} time steps 
  and does not break down when the volume becomes zero. Instead,
  space-time can be extended to a branch preceding the classical
  singularity independently of the matter coupled to the model. For
  large volume the correct semiclassical behavior is
  obtained.
\end{abstract}

\vspace{-6.5cm} 
\begin{flushright}
CGPG--01/2--1  \\
gr-qc/0102069 \\
\end{flushright}
\vspace{5.5cm}


\begin{multicols}{2}

On a macroscopic scale, the gravitational field is successfully described by
general relativity, which is experimentally well tested in the weak
field regime. However, this classical theory must break down in
certain situations where it predicts singularities, i.e.\ boundaries
of space-time which can be reached by observers in finite proper time,
but beyond which an extension of the space-time manifold is impossible
\cite{HawkingEllis}. An outstanding example is the big-bang
singularity appearing in cosmological models. At this point curvature
diverges whence the classical theory completely breaks down and has
to be replaced by a quantum theory of gravity. However, up to now
there is no complete quantum theory of gravity, and so the problem
has been approached by first carrying out a symmetry reduction (by
requiring isotropy and homogeneity) and then quantizing the resulting
minisuperspace models which have only a finite number of degrees of
freedom \cite{DeWitt,Misner}.  In the context of these models, as
yet, there is no definitive resolution of the status of the initial
singularity. Furthermore, generally the methods used in this analysis
can easily miss some key features of the full theory. Indeed, while
it has been speculated for a long time that quantum gravity may lead
to a {\em discrete structure\/} of space and time which could cure
classical singularities, it has not been possible to embody this idea
in standard quantum cosmological models.

By now, there are promising candidates for a quantum theory of
gravity. The results reported in this letter are obtained in the
framework of quantum geometry \cite{Rov:Loops} which {\em does\/} predict
a discrete geometry because, e.g., the spectra of geometric operators
as area and volume are discrete \cite{AreaVol,Area,Vol2}. Although
temporal observables have not been included in the {\em full\/}
theory, it is clear that the space-time structure is very different
from that used in general relativity. But this difference can be
important only at very short scales or in high curvature regimes like
the one close to the classical singularity. This leads to the basic
question raised here: {\em What happens to the classical cosmological
singularity in quantum geometry?}

The first step in our approach is the construction of isotropic states
in full quantum geometry; we {\em first\/} quantize and {\em then\/}
carry out a symmetry reduction. This, however, is not a
straightforward problem because the discrete structure of space,
represented by a graph (spin network) embedded in space, necessarily
breaks any continuous symmetry. But symmetric states can be defined as
generalized states of quantum geometry \cite{SymmRed} which can be
used for a reduction to minisuperspace models \cite{cosmoI}. Note that
this is not a standard symmetry reduction of the classical theory
because symmetric states are interpreted as generalized states in the
{\em full\/} kinematical quantum theory. Only the Hamiltonian
constraint has to be quantized and solved after the reduction. An
immediate and striking consequence is that, in contrast to standard
quantum cosmological models, spatial Riemannian geometry is discrete
leading to a discrete volume spectrum \cite{cosmoII}. Furthermore, in
contrast to standard quantum cosmology, the same techniques as in the
full theory \cite{QSDI} can be used for the quantization of the
reduced Hamiltonian constraint of the cosmological models
\cite{cosmoIII}. This implies another difference, namely that the
evolution equation is not a differential equation in time, but a {\em
difference equation\/} manifesting the discreteness of time
\cite{cosmoIV}.

\paragraph*{Structure of isotropic models.}

According to \cite{SymmRed,cosmoI} states for isotropic models in the
connection representation are distributional states of the full
kinematical quantum theory supported on isotropic connections of the
form $A_a^i=c\Lambda_I^i\omega_a^I$ where $\Lambda_I$ is an internal
$SU(2)$-dreibein and $\omega^I$ are the left-invariant one-forms on
the ``translational'' part of the symmetry group acting on the space
manifold $\Sigma$. The momenta are densitized triads of the form
$E_i^a=p\Lambda_i^IX_I^a$ with left-invariant densitized vector fields
$X_I$ fulfilling $\omega^I(X_J)=\delta^I_J$. Besides gauge freedom,
there are only the two canonically conjugate variables
$\{c,p\}=\kappa\gamma/3$ ($\kappa=8\pi G$ is the gravitational
constant and $\gamma>0$ the Barbero--Immirzi parameter) which have the
physical meaning of extrinsic curvature and square of radius
($a=\sqrt{|p|}$ is the scale factor). The kinematical Hilbert space
${\cal H}_{\rm kin}=L^2(SU(2),{\rm d}\mu_H)$ is the space of functions
of isotropic connections which are square integrable with respect to
Haar measure. Orthonormal gauge invariant states are (see
\cite{cosmoII} for details)
\begin{equation}
 \chi_j=\frac{\sin(j+\case{1}{2})c}{\sin\case{c}{2}}
 \quad,\quad
 \zeta_j=\frac{\cos(j+\case{1}{2})c}{\sin\case{c}{2}}
\end{equation}
for $j\in\case{1}{2}{\Bbb N}_0$ together with $\zeta_{-\frac{1}{2}}=
(\sqrt{2} \sin\case{c}{2})^{-1}$.  These states are eigenstates of the
volume operator $\hat{V}$ with eigenvalues \cite{cosmoII}
\begin{equation}\label{Vj}
 V_j=(\gamma l_{\rm P}^2)^{\frac{3}{2}}
 \sqrt{\case{1}{27}j(j+\case{1}{2})(j+1)}\,.
\end{equation}

Later we will also use a different orthonormal basis of states adapted
to the triad by introducing
\begin{equation}
 |n\rangle:=\frac{\exp(in\case{c}{2})}{\sqrt{2}\sin\case{c}{2}}
 \quad,\quad n\in{\Bbb Z}
\end{equation}
where $n$ represents the eigenvalues of $p$ which determines the
dreibein.  In contrast to $j$, which is always positive and represents
eigenvalues of the square of the scale factor, $n$ can also be
negative. For this it is important that we have not only the character
functions $\chi_j$, but also the additional functions $\zeta_j$. This
concludes the discussion of quantum states.

\paragraph*{The inverse scale factor.}

Classically, the metric of an isotropic spatial slice is given by
$q_{IJ}=a^2\delta_{IJ}=e_I^ie_J^i$ where $e_I^i$ is the co-triad. From
this quantity we can build the expression 
\[
 m_{IJ}:=\frac{q_{IJ}}{\sqrt{\det q}}= \frac{e_I^ie_I^i}{|\det e|}=
 \frac{1}{a}\,\delta_{IJ}
\]
for the inverse scale factor, which we now quantize as a
first application of the previously derived calculus. The co-triad is
not a fundamental variable, but it can be quantized to
$2i(\gamma l_{\rm P}^2)^{-1} h_I[h_I^{-1}, \hat{V}]$ due to the
classical identity $e_a^i=2(\kappa\gamma)^{-1} \{A_a^i,V\}$
\cite{QSDI}.  The expression $\det e$ in the denominator of $m_{IJ}$
can be quantized to the volume operator which then can be absorbed
into the commutators. Such a procedure has already been applied in
\cite{QSDV} in order to quantize matter Hamiltonians which become
densely defined operators, and in the same way we arrive at the {\em
bounded\/} operator
\begin{eqnarray*}
 \hat{m}_{IJ} &=& \case{32}{\gamma^2l_{\rm P}^4}{\rm tr}\left(h_I
 \left[h_I^{-1}, \sqrt{\hat{V}}\right] h_J\left[h_J^{-1},
 \sqrt{\hat{V}}\right]\right)\\
 &=& \case{64}{\gamma^2l_{\rm P}^4}\left(\left(\sqrt{\hat{V}}- \cos\case{c}{2}
   \sqrt{\hat{V}} \cos\case{c}{2} -\sin\case{c}{2} \sqrt{\hat{V}}
   \sin\case{c}{2}\right)^2\right.\\
 & & -\delta_{IJ} \left.\left(\sin\case{c}{2} \sqrt{\hat{V}}
   \cos\case{c}{2}-\cos\case{c}{2} \sqrt{\hat{V}}
   \sin\case{c}{2}\right)^2\right)\,.
\end{eqnarray*}

\begin{figure}
 \centerline{\psfig{figure=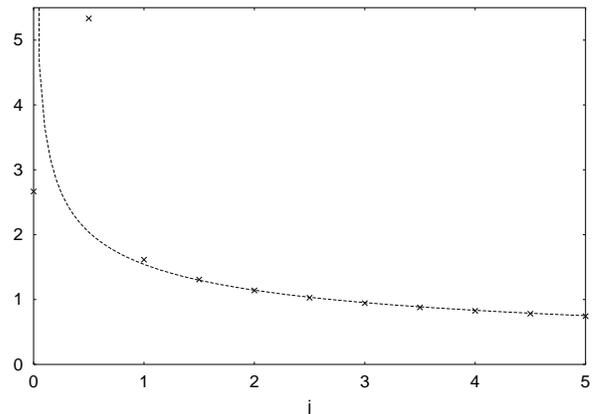,width=3.25in}}
\caption{The classical expectation $V_j^{-\frac{1}{3}}$ (dashed line)
and eigenvalues $m_{II,j}$, $j\geq0$ of the inverse scale factor
($\times$). Contrary to the classical curve, the latter peak at
$j=\frac{1}{2}$ and decrease for $j=0$ and $j=-\frac{1}{2}$
($m_{II,-\frac{1}{2}}=0$ is not shown).}
\label{fig}
\end{figure}

This operator is simultaneously diagonalizable with the volume
operator and has the eigenvalues
\begin{eqnarray}
 m_{IJ,j} &=& \case{16}{\gamma^2l_{\rm P}^4}\left( 4\left(\sqrt{V_j}-
     \case{1}{2} \sqrt{V_{j+\frac{1}{2}}}- \case{1}{2}
     \sqrt{V_{j-\frac{1}{2}}} \right)^2 \right.\nonumber\\
  &&+ \delta_{IJ} \left.\left(\sqrt{V_{j+\frac{1}{2}}}-
    \sqrt{V_{j-\frac{1}{2}}}\right)^2 \right)\\
 &\sim& V^{-\frac{1}{3}}
 \left(\delta_{IJ}+ \case{\gamma^2}{9} \left(\case{1}{256}+ \case{37}{192}
     \delta_{IJ}\right) \case{l_{\rm P}^4}{a^4}\right)
\end{eqnarray}
where in the second step we have assumed that $j$ --- and hence
$V_j$ --- is large. Thus, for large $j$, the leading term is the
classical value $V^{-\frac{1}{3}}\delta_{IJ}$, and the corrections
(which are not necessarily isotropic) are of only the fourth order. We
see that our quantization leading to a bounded operator does not spoil
the classical limit. In fact, the $a^{-1}$-behavior can be observed in
a range which is much larger than expected from the large-$j$
expansion. As Fig.\ \ref{fig} demonstrates, even for $j=1$ are the
eigenvalues very close to the classical expectation, and only the
lowest three eigenvalues show large deviations. But this is already
deeply in the quantum regime, so such deviations are expected and lead
to a finite behavior of the classically diverging $m_{II}$. Note that
the volume operator has the eigenvalue zero (three-fold degenerate),
but even in the corresponding eigenstates is the quantization of the
inverse scale factor perfectly finite. This may be taken as a first
indication for a removal of the classical singularity, although
only at the kinematical level.

\paragraph*{Discrete time evolution.}

Following the basic steps of the quantization in the full theory
\cite{QSDI}, the Hamiltonian constraints for cosmological models can
be quantized with some adaptations to the symmetry \cite{cosmoIII}.
For simplicity we write down here only the key term, the so-called
Euclidean term $H^{({\rm E})}$, of the constraint operator for
spatially flat isotropic models. However, all our qualitative
results remain true for the full constraint and also for isotropic
models with positive curvature. The constraint is of the form
\begin{eqnarray*}
 \hat{H}^{({\rm E})} &=& \case{4i}{\gamma\kappa l_{\rm P}^2}
 \sum_{IJK}\epsilon^{IJK}{\rm tr}(h_Ih_J
 h_I^{-1}h_J^{-1} h_K[h_K^{-1},\hat{V}])\\
 &=& -\case{96i}{\gamma\kappa l_{\rm P}^2}
 \sin^2\case{c}{2}\cos^2\case{c}{2} (\sin\case{c}{2}\hat{V}
 \cos\case{c}{2}- \cos\case{c}{2} \hat{V} \sin\case{c}{2})
\end{eqnarray*}
with action
\begin{eqnarray}
 \hat{H}^{({\mathrm E})}|n\rangle &=& -\case{3}{\gamma\kappa l_{\rm P}^2}
 (V_{|n|/2}- V_{|n|/2-1})\nonumber\\
 && \times (|n+4\rangle-2|n\rangle+|n-4\rangle)\,.
\end{eqnarray}

In order to ``unfreeze dynamics'' and interpret solutions as
``evolving states,'' as usual \cite{kuchar,wsbook} we have to
introduce an internal time which we choose as the dreibein coefficient
$p$. Accordingly, we transform states $|s\rangle$ into an adapted
representation by expanding $|s\rangle= \sum_ns_n|n\rangle$ in
eigenstates $|n\rangle$ of $p$. This will allow us to find an
interpretation of physical states as evolving histories. Furthermore,
discrete geometry imples that eigenvalues of $p$ are discrerte,
whence time evolution is now discrete (see \cite{cosmoIV} for
details). Moreover, since we chose a geometrical quantity as time
which can be negative and is zero for vanishing volume, we will be
able to test the possibility of a quantum evolution through the
classical singularity.

To realize dynamics, we need to extend the model with matter degrees
of freedom which can evolve with this internal time. Matter can be
incorporated by using coefficients $s_n(\phi)$ depending on the matter
field $\phi$ in an appropriate fashion, the details of which is not
important for what follows. The Hamiltonian constraint can then be
written down using a matter Hamiltonian $\hat{H}_{\phi}$ (as in
\cite{QSDV}) which is diagonal in the gravitational degrees of freedom
(and can also contain a cosmological term). The resulting quantum
constraint equation can then be regarded as an evolution in discrete
time:
\begin{eqnarray}\label{WdW}
&&(V_{|n+4|/2}\!-V_{|n+4|/2-1})
 s_{n+4}(\phi)\!-\! 2(V_{|n|/2}\!-V_{|n|/2-1})
 s_{n}(\phi)\nonumber\\
 &&+(V_{|n-4|/2}-V_{|n-4|/2-1}) s_{n-4}(\phi)=
 \case{1}{3}\gamma\kappa l_{\rm P}^2\, \hat{H}_{\phi}\,s_n(\phi)
\end{eqnarray}
($V_j$ are the eigenvalues (\ref{Vj}) of the volume operator with
$V_{-1}=0$) which is a difference equation for the coefficients
$s_n(\phi)$ depending on the discrete label $n$ (our discrete time).

\paragraph*{Fate of the singularity.}

Given initial data $s_n(\phi)$ for some negative $n$, we can use
(\ref{WdW}) in order to determine later values for higher $n$. This,
however, is possible only as long as the highest order coefficient
$V_{|n+4|/2}-V_{|n+4|/2-1}$ is nonzero, which is the
case if and only if $n\not=-4$. So all coefficients for $n<-4$ are
determined by the initial data. However, (\ref{WdW}) does not
determine $s_0$ and instead leads to a {\em consistency condition\/}
for the initial data. So the quantum evolution appears to break-down
just at the classical singularity, i.e.\ at the zero eigenvalue of
$p$. But this is not the case; in fact {\em all\/} $s_n$ for $n>0$ are
determined by (\ref{WdW}) from the initial data. This occurs because
for $n=0$ we have: i) $V_{|n|/2}-V_{|n|/2-1}=0$, and
ii) $\hat{H}_{\phi} s_n(\phi)=0$; thus $s_0$ completely drops out of
the iterative evolution. E.g., $s_4$ is determined solely by $s_{-4}$
because the coefficient of $s_n$ vanishes for $n=0$. So we can evolve
through the singularity and determine all $s_n$ for $n\not=0$. (The
vanishing of $\hat{H}_{\phi} s_0(\phi)$ follows from the quantization
of matter Hamiltonians \cite{QSDV} similarly as described for the
inverse scale factor.)

Of course, in order to determine the complete state we also have to
know $s_0$, but a closer analysis reveals that $s_0$ is fixed from the
outset: The Hamiltonian constraint always has the eigenstate
$s_n=s_0\delta_{n0}$ with zero eigenvalue which is completely
degenerate and not of physical interest. All evolving solutions are
orthogonal to this state and have $s_0=0$ which already fixes the
coefficient $s_0$ left undetermined by using the evolution equation.
We see that the complete state is determined by initial data for
negative $n$, and so there is no singularity in isotropic loop quantum
cosmology. The intuitive picture is as follows: Since for $n<0$ the
volume eigenvalues $V_{(|n|-1)/2}$ decrease with increasing $n$,
there is a contracting branch for negative $n$ leading to a state of
zero volume (in general, $s_{\pm1}\not=0$ and the volume vanishes for
$n=\pm1$ which corresponds to $j=0$) in which the universe bounces off
leading to the expanding branch for positive $n$ which only can be
seen in the classical theory and in standard quantum cosmology. This
conclusion holds true for any kind of matter and cosmological
constant, and is a purely quantum gravitational effect. In particular,
we do not need to introduce matter violating energy conditions and
thereby evade the singularity theorems. However, our result crucially
depends on the factor ordering of the constraint which was chosen as
one of the {\em standard\/} possibilities ordering all triad
components to the right.

\paragraph*{The semiclassical regime.}

We have seen that the classical singularity is removed in loop quantum
cosmology. But we need more for a viable cosmological model, namely we
also need the correct behavior in the semiclassical regime.  Classical
behavior can only be present for large volume and small extrinsic
curvature, i.e.\ if $|n|$ is large, $c$ is small and the wave function
does not vary strongly between successive times $n$ (otherwise the
state would have access to the Planck scale). In this regime we can
interpolate between the discrete labels $n$ and define a wave function
$\psi(a):=s_{n(a)}$, $n(a):=6a^2/\gamma l_{\rm P}^2$ with $a$ ranging
over a continuous range (using $a=\sqrt{|p|}\sim \sqrt{\gamma} l_{\rm
P} \sqrt{|n|/6}$ for large $|n|$ as interpolation points). The
difference operator $\Delta$ then becomes $(\Delta
s)_n:=s_{n+1}-s_{n-1}=\frac{1}{6}\gamma l_{\rm P}^2a^{-1}{\rm
d}\psi/{\rm d}a+O(l_{\rm P}^5/a^5)$ leading to an approximate
constraint operator $\hat{H}^{({\rm E})}\sim-96(i\Delta/2)^2\cdot
a/4\sim -6 \gamma^2 l_{\rm P}^4(-\frac{i}{3}{\rm d}/{\rm d}(a^2))^2a$
for large $a$. This is exactly what one obtains from the classical
constraint $H^{({\rm E})}=-6c^2\sqrt{|p|}$ in standard
quantum cosmology \cite{Kodama} by quantizing $3\hat{c}=-i\gamma
l_{\rm P}^2{\rm d}/{\rm d}p$. In our framework, however, this is only
an approximate equation valid for large scale factors.  For this
equation one can use WKB-techniques in order to derive the correct
classical behavior.

Going to smaller $a$ one has to include more and more corrections in
the expansion of the difference operators and also of the volume
eigenvalues. By doing so one can derive perturbative corrections for
an effective Hamiltonian including higher derivative terms. The closer
we come to the classical singularity, the more corrections we have to
include; and at the singularity we need to know all corrections which,
as we know from our non-perturbative solution, have to add up to yield
the discrete time behavior. So in these models higher order terms
arise from the non-locality in discrete time of the fundamental
theory. But even knowing all perturbative corrections, it would be very
hard to see the correct behavior without knowing the non-perturbative
quantization.

\paragraph*{Quantum Euclidean space.}

In the simplest case, the Euclidean constraint for a spatially flat
model without matter, it is possible to find an explicit solution to
the constraint. The constraint equation is of order eight with one
consistency condition as described above, so one expects seven
independent solutions. But we are interested only in solutions which
have a classical regime in the previous sense, i.e.\ no strong
dependence on $j$ for large $j$. Under this condition one can see that
there is a unique (up to a constant factor) solution
\begin{equation}
 \psi(c)=\sum_j\frac{2j+1}{V_{j+\frac{1}{2}}- V_{j-\frac{1}{2}}}
 \chi_j(c)
\end{equation}
in the connection representation. In standard quantum cosmology the
constraint equation is $\hat{c}^2\sqrt{|\hat{p}|}\xi(c)=0$ with a
solution $\sqrt{|\hat{p}|}\xi(c) =\delta(c)$ which is not unique. In
order to compare the solutions we quantize $a$ by
$\hat{a}\chi_j= 2i (\gamma l_{\rm P}^2)^{-1} (V_{j+\frac{1}{2}}-
V_{j-\frac{1}{2}}) \chi_j$ leading to $\hat{a}\psi\propto
\sum_j(2j+1)\chi_j$ which in fact is the delta function on the
configuration space $SU(2)$.  Therefore, we have a unique solution
which incorporates the characterization of Euclidean space to have
vanishing extrinsic curvature of its flat spatial slices.

\paragraph*{Conclusions.}

We have shown in this paper that canonical quantum gravity is
well-suited to analyze the behavior close to the classical
singularity. For this, it is important to use only techniques which
are applicable in the {\em full\/} theory. This leads to a discrete
structure of space and time which cannot be seen in standard quantum
cosmology. In our framework, the standard quantum osmological
description arises only as a limit for large volume where the
discreteness is unimportant. For small volume, quantum geometry leads
to new effects which are responsible for the removal of the classical
singularity. In contrast to earlier attempts this is {\em not\/} achieved
by introducing matter which violates energy conditions; it is a pure
quantum gravity effect. It also does not avoid the zero volume state
present in the classical singularity because in general the wave
function is not orthogonal to states with zero volume
eigenvalue. Nevertheless there is no sign of a singularity because in
quantum geometry it is possible to have vanishing volume but
non-diverging inverse scale factor, which in isotropic models dictates
all curvature blow-ups. Besides removing the singularity, the fact
that an evolution through a state of zero volume is possible without
problems could lead to topology change in quantum gravity.
Technically, the removal of the singularity is achieved by using
Thiemann's strategy \cite{QSDI} of absorbing inverse powers of
$\hat{V}$ into a Poisson bracket which also lead to densely defined
matter Hamiltonians \cite{QSDV}. So it is the same mechanism which
regularizes ultraviolet divergences in matter field theories and which
removes the classical cosmological singularity. We have also seen that
non-perturbative effects are solely responsible for this behavior and
a purely perturbative analysis could not lead to these conclusions.

\paragraph*{Acknowledgements.}

The author is grateful to A.\ Ashtekar for suggesting a study of the
implications of discrete volume and time close to the classical
singularity, for discussions and for a critical reading of the
manuscript, and to H.\ Kastrup for comments.  This work was supported
in part by NSF grant INT9722514 and the Eberly research funds of Penn
State.

\end{multicols}



\end{document}